\documentclass[12pt]{iopart}
\usepackage{graphicx}

\begin{document}
\title[Current-voltage characteristics of tunable F-Si-F channels]
      {Current-voltage characteristics of tunable ferromagnet-silicon-ferromagnet
       channels in the spin blockade regime}

\author{D V Khomitsky}

\address{Department of Physics, University of Nizhny Novgorod,
         Gagarin Avenue 23, 603950 Nizhny Novgorod, Russian Federation}
\ead{khomitsky@phys.unn.ru}

\begin{abstract}
The steady current-voltage characteristics of ferromagnet-silicon-ferromagnet channels with tunable emitter
and collector polarizations are investigated in the presence of spin blockade generalizing the model developed
by Pershin Yu V and Di Ventra M (2008 {\it Phys. Rev.} B {\bf 77} 073301). The dependence of the critical
current on both collector and emitter polarizations is obtained analytically. It is found that the current
amplitude in the channel can be effectively tuned by varying the difference between the collector and emitter
ferromagnet polarizations which allows to perform the magnetic manipulation of the electrical current in wide
class of both n- and p-doped, low- and high-Ohmic semiconductor channels coupled to ferromagnetic leads.
\end{abstract}


\submitto{\JPCM}

\maketitle

\section{Introduction}

The progress of spintronics and physics of heterostructures which can be observed
during the last years \cite{Awschalom,Zutic} is focused on various physical phenomena,
and one of them which attracts a considerable attention is the spin-dependent
transport through semiconductor/spin-polarized junctions
\cite{Kawakami,Zutic02,Epstein,Yu,Albrecht02,Albrecht03,Yu03,Stephens,Crooker,Dery,Pershin07,Pershin,Pershin08}.
The physics of carrier polarization and its influence on transport in composite structures
such as semiconductor/ferromagnet has been studied both theoretically
\cite{Zutic02,Yu,Albrecht02,Albrecht03,Yu03,Dery,Pershin07,Pershin,Pershin08}
and experimentally \cite{Kawakami,Epstein,Stephens,Crooker}.
One of the models describing the spin-resolved carrier concentrations and currents at the junction
is the two-component drift-diffusion model \cite{Yu,Pershin07,Pershin} which predicted
highly nonlinear and saturating current-voltage dependence at a single semiconductor/feromagnet
junction due to the effect of spin blockade \cite{Pershin07,Pershin}.
In this model the detailed structure of the charge and current distribution at
the junction area \cite{Zutic02,Yu} as well as the Schottky barriers \cite{Albrecht02,Albrecht03},
the charge redistribution effects \cite{Yu03}, and the bound states \cite{Dery} are not taken
into consideration. Still, the qualitative and distinguishable behaviour of current saturation due
to the effect of spin blockade is reliably predicted under various system parameters such
as the junction/semiconductor resistance ratio. The spin blockade regime arises from the spatial
distribution of the spin-minority carriers which cannot enter the ferromagnet region and form
a cloud near the junction which growth prevents the further increase of spin-majority carrier transport
if the current exceeds a threshold value called the critical current. Further studies have shown
the importance and promising applications of this effect also for non-stationary phenomena such
as spin memory effects \cite{Pershin08}. The models described above were applied mainly to GaAs-based
semiconductor channels, but is is known that the silicon-based structures are also of big interest
for spintronics due to the dominating place of silicon in currently available electronic technologies.
More, the technologies of fabricating the silicon/ferromagnet structures such as Si/Si:Mn formed
on a basis of diluted magnetic semiconductors have been intensively developed during the last few years
\cite{Demidov06,Demidov09} which makes their future applications in spintronics
promising and creates certain questions about the phenomena described above.
Is there a spin blockade regime in a silicon/ferromagnet junction at specific values of applied voltage,
carrier mobility and concentration ? If so, what is the critical current density and how it depends on
the silicon and ferromagnet parameters such as the carrier polarization in ferromagnets and
the conductivity of the semiconductor channel? How deep can we modulate the current in the channel by
manipulating the polarization of emitter or collector ferromagnets relative to each other?
In the present manuscript we study these problems in the framework of a simple but effective model
of transport in the spin blockade regime \cite{Pershin07,Pershin} which we generalize for the case
of arbitrary carrier polarizations in the emitting and collecting ferromagnetic regions of
the channel as well as for wide range of low- and high-Ohmic n-doped and p-doped silicon samples.
It is found that the current can be deeply modulated by changing the spin alignment
in the emitter and/or collector ferromagnet since the critical current density
is very sensitive to it. We find the analytical expression for the critical current
density and calculate the current-voltage dependencies for various combinations of
the channel/contact resistance ratios, as well as for n- and p-type of doping with both high
and low concentrations. The manuscript is organized as follows: in Section 2 we
derive a model generalizing the description of the spin blockade regime for the two-ferromagnet
channel with arbitrary polarizations in the emitter and collector ferromagnets and discuss
the properties of the critical current density, in Section 3 we plot and discuss the current-voltage
characteristics for various combinations of system parameters, and the conclusions are given in Section 4.

\section{The model}

\begin{figure}
  \centering
  \includegraphics[width=85mm]{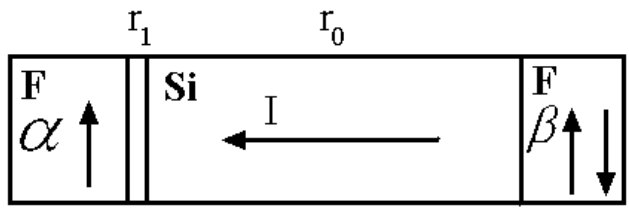}
  \caption{Schematic view of a ferromagnetic (F) - silicon (Si) - ferromagnetic (F) channel
           with variable polarization degrees $\alpha$ and $\beta$ in collector and emitter ferromagnets.
           The channel and contact zero-bias resistances are $r_0$ and $r_1$, and the direction of the carrier motion
           $I$ is shown (for the electrons the actual current is $-I$).
           The spin blockade may occur in the contact region at the collector (left F) if the polarizations
           $\alpha$ and $\beta$ significantly differ from each other, which enhances the contact resistance from
           the zero-bias value $r_1$ as long as the current builds up.}
  \label{fsif}
\end{figure}

The schematic view of the ferromagnet-silicon-ferromagnet channel is shown in
Figure \ref{fsif}. The collector ferromagnet with the junction resistance $r_1$ is separated from
the emitter ferromagnet by a bulk silicon channel with length $L$ which we consider as exceeding
the spin diffusion length $l_s$ given by \cite{Yu,Pershin07,Pershin}

\begin{equation}
l_s=\frac{2D}{\mu E +\sqrt{\mu^2 E^2 + 4D/\tau_s}}
\label{ls}
\end{equation}

where $D$ and $\mu$ are the carrier diffusion coefficient and drift mobility, respectively, $E$ is the electric field
inside the channel, and $\tau \sim 10$ ns is the typical spin relaxation time \cite{Zutic}. One can
see from (\ref{ls}) that $l_s$ is maximal at zero electric field when $l_s=\sqrt{D \tau_s}$
and when the diffusion coefficient and spin relaxation time are big, or, since $D$ is coupled to the mobility via
the Einstein relation $D=\mu k_B T/e$ \cite{Smith}, it is clear that the spin diffusion length grows with the mobility.
Since the mobility in the bulk silicon is typically lower than the one for GaAs \cite{Sze}, for the given parameters
of the silicon the condition $L>>l_s$ is already fulfilled if $L \ge 10$ mkm which is a reasonable
channel length of bulk semiconductor structures. Hence, from the point view where the spin-resolved
concentration decay length is considered, the distance between emitter and collector ferromagnets here can
be taken as infinite which simplifies the boundary conditions. The key parameters of our model are the variable
polarization degrees $\alpha$ and $\beta$ in the collector and emitter ferromagnets describing the state of their
non-ideality as well as the chosen direction of polarization for the majority of carriers.
For example, the pair $\alpha=1$, $\beta=1/2$ corresponds to the previously investigated case \cite{Pershin}
where the carriers are fully polarized in the collector and fully unpolarized in the emitter.
Correspondingly, the choice  $\alpha=0.8$, $\beta=0.7$ describes the situation when
the collector and emitter are highly polarized in the same direction while the values $\alpha=0.8$,
$\beta=0.3$ describe the opposite polarization at the emitter.
We adopt the labeling $j_{1(2)}$ for the spin-resolved current with spins in ferromagnets aligned up
(down). If $j$ is the total current density, than the spin-resolved current at the collector
$j_1=\alpha j$ which means that $\alpha=1(0)$ corresponds to an ideal ferromagnet
where magnetic moments are all aligned up (down), and $\alpha=1/2$ describes the
unpolarized current. The same labeling is adopted for the carrier polarization generated by the emitter
ferromagnet with polarization $\beta$. The main purpose of the present paper is
to find out how deep is the influence of the various $\alpha$ and $\beta$ differences reflecting the polarization
switch on the total electrical current in the channel. We include in our model the channel and contact zero-bias
resistances $r_0$ and $r_1$ (see Figure \ref{fsif}) with variable ratio, and the direction of the carrier motion
is labeled as $I$ (for the electrons the actual current is $-I$). The spin blockade may occur in the contact region
at the collector (left F in Figure \ref{fsif}) if the polarizations $\alpha$ and $\beta$ significantly differ from
each other, which enhances the contact resistance from the zero-bias value $r_1$ as long as the current builds up.

The model for the spin-resolved current densities $j_{1,2}$ and spin-resolved concentrations $n_1+n_2=N$ where
$N$ is the total carrier concentration is well-known \cite{Yu,Pershin07,Pershin} and consists of the continuity equation

\begin{equation}
e\frac{\partial n_{1,2}}{\partial t}={\rm div}{\vec j}_{1,2}
+\frac{e}{2\tau_s}(n_{2,1}-n_{1,2})
\label{ce}
\end{equation}

and the equation for the spin-resolved current

\begin{equation}
{\vec j}_{1,2}=en_{1,2}\mu E + eD \nabla n_{1,2}.
\label{js}
\end{equation}

These equations should be accompanied by the boundary conditions at the collector
contact $x=0$ and at the other boundary of the channel $x_{\rm{max}}=L >> l_s$ which is in our
problem can be considered as $x=\infty$. In this case the steady-state solution of (\ref{ce}) exists
\cite{Pershin07,Pershin} which has the form of two exponents decaying to their concentrations
defined by the boundary conditions at the emitter. In our model the emitter ferromagnet is
described by the arbitrary polarization $\beta$ which generalizes the unpolarized case $\beta=1/2$ considered
previously \cite{Pershin07,Pershin}, so

\begin{eqnarray}
n_1(\infty)=\beta N \\
n_2(\infty)=(1-\beta) N.
\label{nem}
\end{eqnarray}

The steady-state solution of (\ref{ce}) satisfying the boundary condition (\ref{nem}) at the emitter
and the normalizing condition $n_1+n_2=N$ has the form

\begin{eqnarray}
n_1(x)=\beta N - A e^{-\lambda x} \\
n_2(x)=(1-\beta) N + A e^{-\lambda x},
\label{nx}
\end{eqnarray}

where $\lambda=1/l_s$ is the inverse decay length defined in (\ref{ls}).
The parameter $A$ should be determined from the boundary condition at the collector when
$x=0$. If the collector is a non-ideal ferromagnet with the polarization $\alpha$,
then the spin-resolved current at the collector

\begin{equation}
j_1(0)=\alpha j,
\label{jcoll}
\end{equation}

and the other boundary condition at the emitter $j_1(\infty)=\beta j$ is satisfied
automatically. The total current density $j$ is related to the electric field in the
channel via usual relation $j=e\mu N E$ which allows to construct a closed equation for the current density and
applied voltage. From (\ref{jcoll}) it follows that the conductivity of the collector junction is proportional to
the concentration of the majority carriers, and at the absence of the current
the junction resistance has a predetermined value $r_1$ while the silicon channel with length $L$ and cross-section $S$
is described by the resistance defined in a usual way by the carrier mobility, concentration and  geometric dimensions
as $r_0=L/(e\mu N S)$. After a simple algebra we obtain the Ohmic law $I(r_0+r_1(I))=V$ which reads in our case
similar to the one obtained previously \cite{Pershin} for a non-polarized emitter
with $\beta=1/2$,

\begin{equation}
I\left(r_0+r_1\frac{1}{1-\frac{2}{\sqrt{1+8\left[\frac{j_c S}{I}\right]^2}-1}}\right)=V,
\label{cv}
\end{equation}

but in our generalized model the parameter $j_c$ defined as the critical current density \cite{Pershin}
depends on both the collector and emitter ferromagnet polarizations $\alpha$ and $\beta$:

\begin{equation}
j_c=j_c^{0}\sqrt{\frac{2}{\left(\frac{\alpha-\beta/2}{\beta} \right)^2-\frac{1}{4}}}
\label{jc}
\end{equation}

where

\begin{equation}
j_c^{0}=eN\sqrt{\frac{D}{2\tau_s}}
\label{jc0}
\end{equation}

is the critical current density for the case of fully polarized collector ferromagnet
with $\alpha=1$ and fully unpolarized emitter with $\beta=1/2$ \cite{Pershin}. Indeed,
for $\alpha=1$ and $\beta=1/2$ the value of $j_c$ from (\ref{jc}) is equal to
$j_c^{(0)}$.

\begin{figure}
  \centering
  \includegraphics[width=85mm]{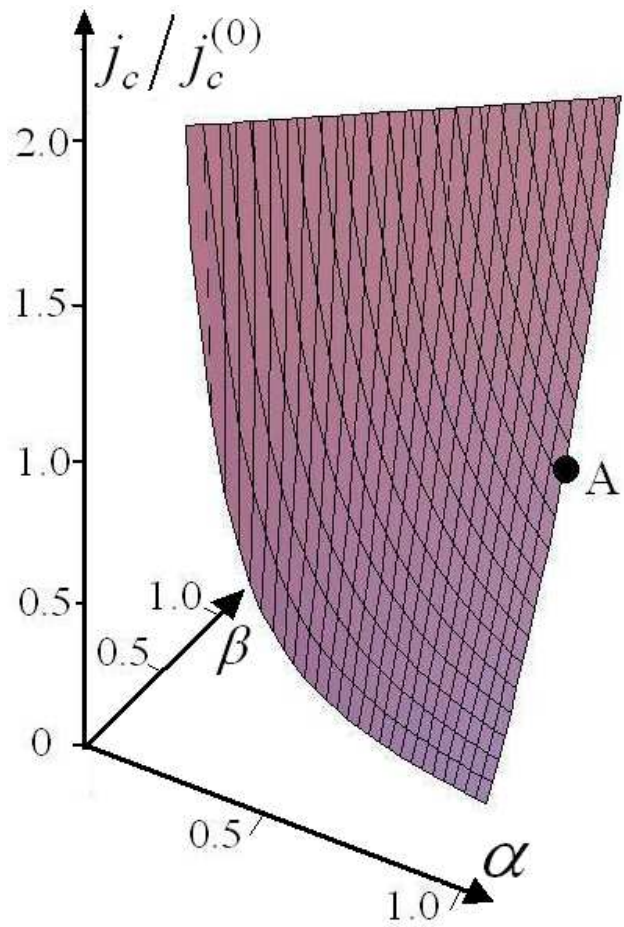}
  \caption{(Colour online) Dependence of critical current density $j_c$ measured in units of $j_c^{(0)}$ on
           the polarizations $\alpha$ and $\beta$ in collector and emitter ferromagnets. The case of fully polarized
           carriers in the collector with $\alpha=1$ and fully unpolarized electrons in the emitter $\beta=1/2$ considered
           in \cite{Pershin} is shown by a reference point A when $j_c=j_c^{(0)}$. The magnitude of $j_c$ diverges along
           the line $\alpha=\beta$ which means that in the case of a perfect coincidence of the spin alignment in
           the emitter and collector the critical current can never be reached and there is no spin blockade in this limit.
           In the opposite limit of maximum difference between $\alpha$ and $\beta$ the value of $j_c$ is considerably
           smaller than $j_c^{(0)}$, so the spin blockade here occurs at lower current densities.}
  \label{fjc}
\end{figure}

The main property of equation (\ref{cv}) determining the
current-voltage characteristic is the current saturation occurring at current
densities comparable to $j_c$. Thus, it is of interest to investigate the critical
current density (\ref{jc}) in more details with respect to the tunable polarization parameters
$\alpha$ and $\beta$ of the collector and emitter ferromagnets. In Figure \ref{fjc}
the three-dimensional plot $j_c=j_c(\alpha,\beta)$ is shown for the critical current density (\ref{jc}) measured
in units of $j_c^{(0)}$. The case of fully polarized carriers in the collector with $\alpha=1$ and fully
unpolarized electrons in the emitter $\beta=1/2$ considered in \cite{Pershin} is shown by a reference point A when
$j_c=j_c^{(0)}$. It is clear from (\ref{jc}) that $j_c$ diverges along the line $\alpha=\beta$ which means that
in the case of a perfect coincidence of the spin alignment in the emitter and collector the critical current
can never be reached and there is no spin blockade in this limit. In the opposite limit of maximum difference between
$\alpha$ and $\beta$ the value of $j_c$ is considerably smaller than $j_c^{(0)}$, so the spin blockade here occurs
at lower current densities. The observed large variations of the critical current
density are important for the desired high degree of control on the electrical current
flowing through the ferromagnet-silicon-ferromagnet channel which can be achieved by
variations of at least one ferromagnet polarization, say the emitter polarization $\beta$.
Below we shall see that these expectations are confirmed by the current-voltage
dependencies in both n-doped and p-doped channels with both low- and high-Ohmic
resistance.

\section{Current-voltage characteristics}

\begin{figure}
  \centering
  \includegraphics[width=85mm]{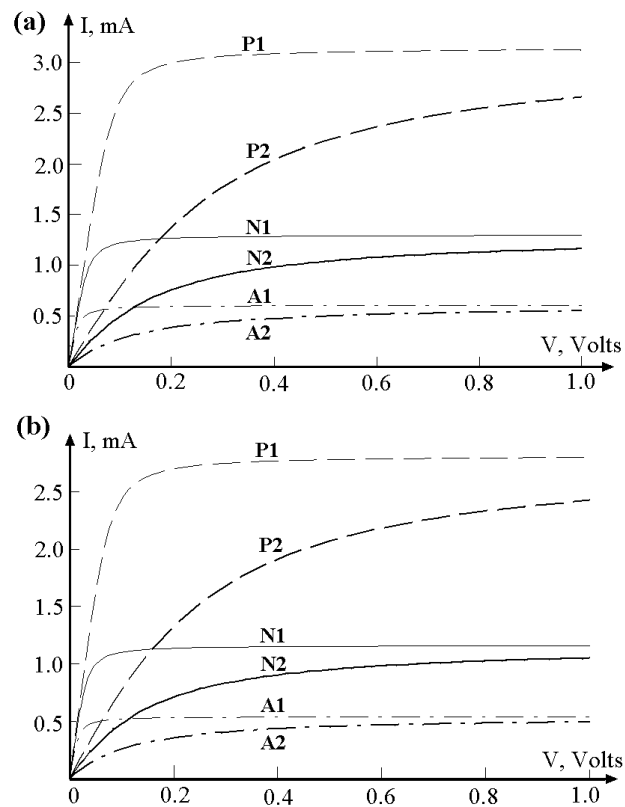}
  \caption{Current-voltage characteristics for a F-Si-F channel with $L=10$ mkm and $S=5$ $\rm{mkm^2}$ shown
          in Figure \ref{fsif} with (a) n-doped and (b) p-doped
          silicon with high carrier concentration $N=10^{19}$ $\rm{cm^{-3}}$. On each plot the three families of curves
          correspond to three positions of the "spin valve" emitter while the carriers at the collector are always highly
          polarized with $\alpha=0.8$. Each family has two curves labeled 1 and 2 (bold) which corresponds to the low and
          high ratios of the junction and channel resistances $r_1/r_0=1/5$ and $r_1/r_0=5/1$. The solid curves N1, N2
          describe the case of non-polarized carriers at the emitter with $\beta=0.5$ which was considered
          in \cite{Pershin}, the dashed curves P1, P2 are for the emitter polarization $\beta=0.7$ which is close
          to the collector polarization $\alpha=0.8$, and the dash-dotted curves A1,A2 are for the emitter polarization
          $\beta=0.3$ which differs significantly from the one of the collector. For both type of carriers and for both
          low and high ratios of the contact/junction resistance the current saturation is observed being the indication
          of the spin blockade regime. The switch between different emitter polarizations leads to strong modulations
          of the current saturation amplitude, creating the possibility of the magnetization control on the current.}
  \label{fvah1}
\end{figure}

\begin{figure}
  \centering
  \includegraphics[width=85mm]{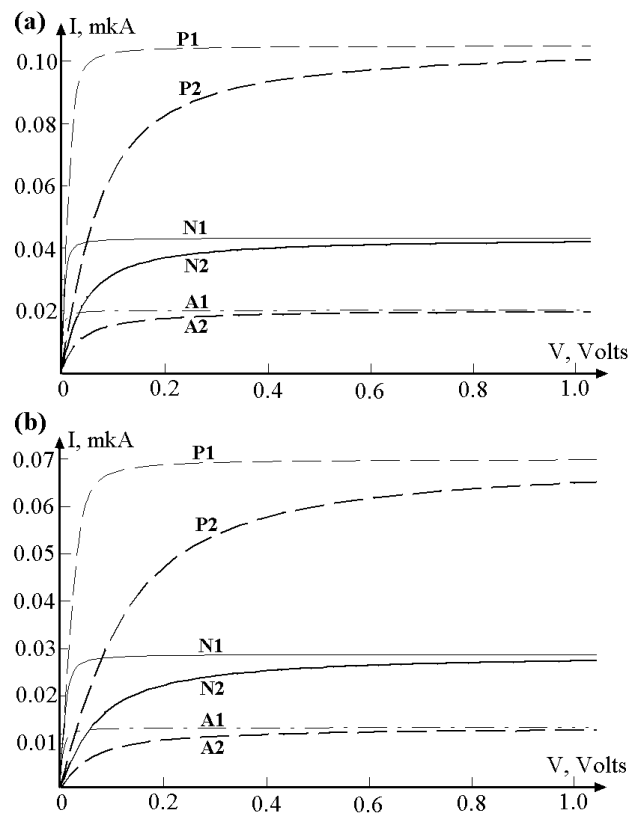}
  \caption{Same as in Figure \ref{fvah1} but for the (a) n-doped and (b) p-doped silicon channels with low carrier
           concentration $N=10^{14}$ $\rm{cm^{-3}}$. The labeling of curves is the same as in Figure \ref{fvah1},
           and the maximum achievable currents are lower due to significantly smaller carrier concentration which
           decreases the conductivity. The same conclusions about the spin blockade manifestation and polarization
           control of the current amplitudes mentioned in the caption for Figure \ref{fvah1} can be applied here for
           the case of low carrier concentrations.}
  \label{fvah2}
\end{figure}

The current-voltage characteristics for n-doped and p-doped silicon channels with high
and low carrier concentration are shown in Figures \ref{fvah1} and \ref{fvah2},
respectively.

The data for the room temperature drift mobility of electrons and holes in the bulk silicon
doped with various concentrations was taken from the standard reference plots \cite{Sze},
allowing to estimate both channel resistance for a bulk silicon sample with $L=10$ mkm and $S=5$ $\rm{mkm^2}$
and the reference critical current density $j_c^{(0)}$ from (\ref{jc0}). In Figure \ref{fvah1}
the results are shown for the silicon channel with high carrier concentration
$N=10^{19}$ $\rm{cm^{-3}}$ where the reference critical current density $j_c^{(0)}$ is within the range of
$160\ldots 180$ $\rm{mkA/mkm^2}$, and the channel resistance $r_0=25\ldots 30$ Ohm. On each plot the three
families of curves correspond to three positions of the "spin valve" emitter while the carriers at the collector
are always highly polarized with $\alpha=0.8$. Each family has two curves labeled 1 and 2 (bold) which corresponds
to the low and high ratios of the junction and channel resistances $r_1/r_0=1/5$ and $r_1/r_0=5/1$.
The solid curves N1, N2 describe the case of non-polarized carriers at the emitter with $\beta=0.5$ which was considered
in \cite{Pershin}, the dashed curves P1, P2 are for the emitter polarization $\beta=0.7$ which is close
to the collector polarization $\alpha=0.8$, and the dash-dotted curves A1,A2 are for the emitter polarization
$\beta=0.3$ which differs significantly from the one of the collector. For both type of carriers and for both
low and high ratios of the contact/junction resistance ratio the current saturation is observed being the indication
of the spin blockade regime. The switch between different emitter polarizations leads to strong modulations
of the current saturation amplitude, creating the possibility of the magnetization control on the current.

To compare the results for low-Ohmic samples with the high-Ohmic ones, we present in Figure \ref{fvah2}
the current-voltage characteristics for the silicon channel doped with low carrier concentration
$N=10^{14}$ $\rm{cm^{-3}}$ where the reference critical current density is much lower and is within
the range of $0.002\ldots 0.003$ $\rm{mkA/mkm^2}$. The channel resistance is correspondingly much
higher, and we consider $r_0=200\ldots 500$ kOhm for n-doped and p-doped samples, respectively.
The labeling of all curves is the same as in Figure \ref{fvah1}, and the maximum achievable currents are much lower
due to significantly smaller carrier concentration which decreases the conductivity.
One can draw the same conclusions about the spin blockade manifestation and polarization
control of the current amplitudes mentioned above which obviously can be applied also
for the case of low carrier concentrations. We can conclude that the main goal of the present model
which is the achievement of a deep current modulation by the magnetization switch at the emitter
can be reached in both low- and highly-doped silicon samples, with both n-type and p-type doping
and with both low and high ratios of the junction/channel resistances. Hence, the proposed model
of the tunable current-voltage characteristics in a ferromagnet-silicon-ferromagnet channel seems
to be applicable to a rather wide range of ferromagnet-semiconductor structures.

\section{Conclusions}
We have studied the steady-state current-voltage characteristics of ferromagnet-silicon-ferromagnet channels with
long bulk silicon sample having the length exceeding the spin diffusion length. The current behaviour was investigated
in the presence of spin blockade regime at the collector junction, and the dependence of the critical current
on both collector and emitter polarizations has been obtained analytically. It was found that the current amplitude
can be effectively tuned by varying the difference between the collector and emitter ferromagnet polarizations which
allows to perform the magnetic manipulation on the electrical current in wide class of both n- and p-doped, low- and
high-Ohmic semiconductor channels coupled to ferromagnetic leads.

\ack
The author is grateful to E.S. Demidov for many helpful discussions. The work was
supported by the Ministry of Education and Science RF under the Program "Development of the Higher School
Research Potential" (Grant No. 2.1.1/2833).

\end{document}